\documentclass[aps,prb,twocolumn,showpacs,preprintnumbers,amsmath,amssymb,superscriptaddress]{revtex4}
\usepackage{dcolumn}
\usepackage{bm}
\usepackage{graphicx}
\usepackage{times}
\usepackage{epstopdf}
\usepackage{color}
\usepackage{tikz}
\usetikzlibrary{decorations.markings}
\usepackage[subrefformat=parens,labelformat=parens,caption=false]{subfig}
\tikzset{->-/.style={decoration={markings, mark=at position 0.5 with {\arrow{>}}},postaction={decorate}}}
\tikzset{-<-/.style={decoration={markings, mark=at position 0.5 with {\arrow{<}}},postaction={decorate}}}
\usepackage[pdftex,colorlinks,bookmarks=true]{hyperref}
\usepackage[capitalize]{cleveref} 
\begin{document}

\definecolor{Red}{rgb}{1,0,0}
\definecolor{Blu}{rgb}{0,0,01}
\definecolor{Green}{rgb}{0,1,0}
\newcommand{\red}{\color{Red}}
\newcommand{\blu}{\color{Blu}}
\newcommand{\green}{\color{Green}}
\newcommand{\bv}[1]{\boldsymbol{#1}}
\newcommand{\bvh}[1]{\hat{\bv{#1}}}

\title{Current-loops, phase transitions, and the Higgs mechanism in
Josephson-coupled multi-component superconductors}

\author{Peder Notto Galteland}
\affiliation{Department of Physics, NTNU, Norwegian University of
Science and Technology, N-7491 Trondheim, Norway}
\author{Asle Sudb\o}
\affiliation{Department of Physics, NTNU, Norwegian University of
Science and Technology, N-7491 Trondheim, Norway}

\date{\today}
\begin{abstract}
  The $N$-component London $\mathrm{U}(1)$ superconductor is expressed in terms of integer-valued
  supercurrents. We show that the inclusion of inter-band Josephson couplings introduces
	monopoles in the current fields, which convert the phase transitions of the charge-neutral
	sector to crossovers. The monopoles only couple to the neutral sector, and leave the phase
  transition of the charged sector intact. The remnant non-critical fluctuations
	in the neutral sector influence the one remaining phase transition in the charged sector, and
	may alter this phase transition from a $3DXY$ inverted phase transition into a first-order
	phase transition depending on what the values of the gauge-charge and the inter-component
	Josephson coupling are. This preemptive effect becomes more pronounced with increasing number
	of components $N$, since the number of charge-neutral fluctuating modes that can influence
	the charged sector increases with $N$. We also calculate the gauge-field
  correlator, and by extension the Higgs mass, in terms of current-current correlators. We
	show that the onset of the Higgs-mass of the photon (Meissner-effect) is given in terms
	of a current-loop blowout associated with going into the superconducting state as the
	temperature of the system is lowered.
\end{abstract}
\maketitle

\section{Introduction}

Models with multiple $\mathrm{U}(1)$ condensates coupled by a vector potential are relevant to a
variety of condensed matter systems. The number of possible interactions between the individual
condensates make the models much more complex than single-band systems. Multiple, individually
conserved condensates are applicable to systems of low temperature atoms, such as hydrogen under
extreme
pressures\cite{Jaffe1981,Oliva1984,Moulopoulos1991,Moulopoulos1999,Ashcroft2003,Ashcroft2004}
and as effective models of easy-plane quantum anti-ferromagnets\cite{Senthil2004,Motrunich2004}.
Superconductors with multiple superconducting bands, such as
$\text{MgB}_2$\cite{Bouquet2001,Liu2001,Szabo2001} and iron pnictides\cite{YoichiKamihara2008} may
also be described by a model of multiple $\mathrm{U}(1)$ condensates, but in these systems the
individual condensates are not conserved. Inter-band Josephson couplings must always be included,
as they cannot  \textit{a priori} be excluded on symmetry grounds.

Ginzburg-Landau models of $N$-component superconductors in the London limit host a rich variety of
interesting phenomena\cite{Babaev2004,Smorgrav2005,Smorgrav2005a,Kuklov2006}. Each condensate supports topological vortex
line defects, which represent disorder in the condensate ordering field. When the condensates are
coupled through a gauge field, the vortices carry magnetic flux quanta, and may be bound into
composite vortices with $\pm 2\pi$ phase windings in multiple condensates\cite{Smiseth2005}.  It
turns out that this gives rise to composite superfluid modes that do not couple to the gauge field,
even though their constituent vortices interact via the gauge field. In addition to the superfluid
modes, there will be a single charged mode which is coupled by the gauge field. This causes the
$N$-component model without Josephson interactions to have $N-1$ superfluid phase transitions and a
single superconducting phase transition\cite{Smiseth2005}. For certain values of the gauge charge
these transitions will interfere in a non-trivial way, causing the transitions to merge in a single
first-order transition\cite{Kragset2006, Herland2010}.

\begin{figure}
\includegraphics[width=\columnwidth]{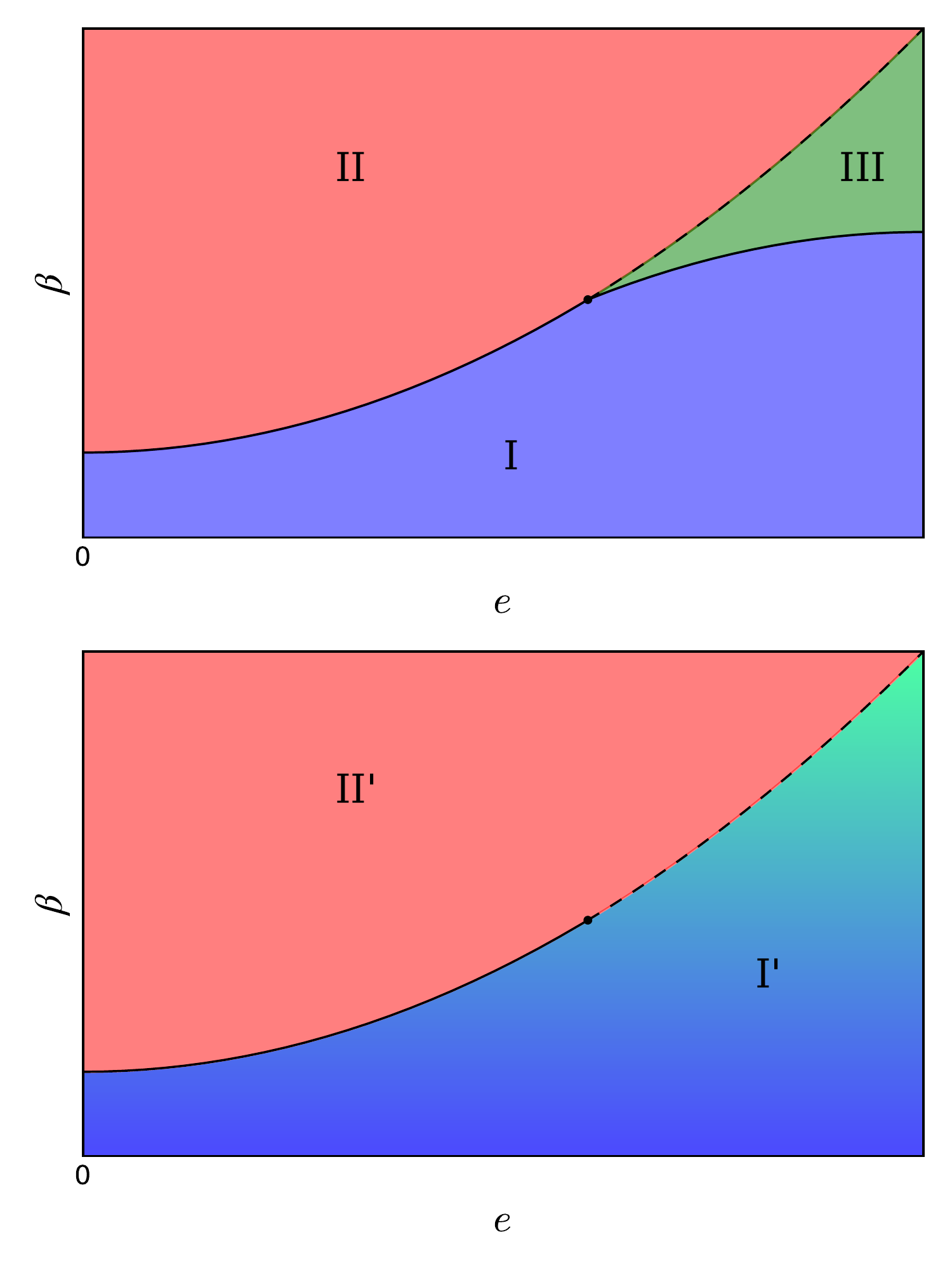}
\caption{A schematic phase diagram for the model with $N=2$. The top panel shows the
case $\lambda=0$, while the lower panel shows the case with $\lambda>0$.
Top panel, $\lambda=0$: Phase I is the fully symmetric normal phase with no superfluidity and no superconductivity.
Phase III is the phase with no superconductivity, but non-zero
superfluid stiffness in the neutral mode (metallic superfluid). Phase II is the low-temperature fully ordered state
with finite Higgs mass in the charged sector and finite superfluid density in the neutral sector, a superconducting
superfluid. The solid line separating phase I from phase II is a first-order phase transition line. The dotted line
separating phase II from phase III is a critical line in the inverted $3DXY$ universality class. The solid line
  separating phase I from phase III is a critical line in the $3DXY$ universality class. {Along
  the line $e=0$, we recover two uncoupled $3DXY$ models, and the phase transition will be two
  superimposed independent phase transitions in the $3DXY$ universality class.} Bottom panel, $\lambda > 0$:
Phase I' is the high-temperature phase with no superconductivity. The entire phase is analytically connected with only
a crossover regime separating the high-temperature phase from the lower-temperature phase. There is no spontaneous
symmetry breaking in the neutral sector, since the Josephson coupling effectively acts as an explicit symmetry-breaking
term in this sector, analogous to a magnetic field coupling linearly to $XY$ spins. Phase II' is the low-temperature
superconducting state. The solid part of the line separating phase I' from phase II' is a first-order phase transition
line. The dotted part is a critical line in the inverted $3DXY$ universality class. Both for $\lambda=0$ and $\lambda> 0$,
the line separating the superconducting states (II and II') from the non-superconducting state changes character from a
first-order phase transition (solid line) to a second-order phase-transition (dotted line) as via a tricritical point.
The $3DXY$ critical line separating phase I from phase III for $\lambda =0$, is converted to a crossover line in
  phase I' for $\lambda >0$. { Along the $e=0$ line, the system is described by two neutral sectors coupled by an inter-component Josephson coupling, such that the global
		$U(1) \times U(1)$ symmetry is reduced to a global $U(1)$-symmetry. Therefore, 
		the phase transition  reverts to a single $3DXY$ transition.}}
\label{fig:PD}
\end{figure}

The question of the nature of the phase transitions present in Josephson-coupled multiband
superconductors is of considerable interest. Symmetry arguments dictate that the inclusion of the
Josephson coupling breaks the $\left[\mathrm{U}(1)\right]^n$ symmetry down to $\mathrm{U}(1)$, at
any strength. The Josephson term locks the superfluid modes so that the phase transition in the
neutral sector is replaced by a crossover\cite{Smiseth2005}, while the phase transition in the
gauge-coupled sector is expected to remain. If this transition remains continuous, it is expected to
be in the inverted 3D$XY$ universality class\cite{Smiseth2005}. A recent study has observed a first
order transition in this model for weak Josephson coupling\cite{Sellin2016}, suggesting a subtle
interplay between the two length scales dictated by the Josephson length and the magnetic field
penetration depth.  A schematic phase diagram is shown in \cref{fig:PD} for the two-component
case. This is based on arguments provided in this work, and supports the numerical results obtained
in recent numerical studies\cite{Sellin2016}. Also of note is multiband superconductors with
frustrated inter-band couplings, which is $\mathrm{U}(1)\times\mathrm{Z}_2$ symmetric. These systems
have been shown to have a single first-order transition in three dimensions from a symmetric state
into a state that breaks both $\mathrm{U}(1)$ and $\mathrm{Z}_2$ symmetry for weak values of the
gauge field coupling. For stronger values of the charge, the transitions
split\cite{Bojesen2014,Bojesen2015}.

In this paper we present an alternate approach to the multiband superconductor which has
certain advantages over standard formulations, allowing further analytical
insights to be made.
In particular, we are able reconcile the different results for the character of the phase transition in the charged sector found in Refs. \onlinecite{Smiseth2005} and
\onlinecite{Sellin2016} in the presence of interband Josephson-couplings. By applying a
character expansion\cite{Janke1986,Kleinert1989} to the action, we replace the phases of the
order parameter with integer-current fields. These currents are the actual supercurrents of the
model. \cref{sec:model} presents the details and basic
properties of the multiband superconductor in the London limit. In \cref{sec:bexpJ0} we present the
character expansion, apply it to the model with no Josephson coupling, and compare the resulting
representation to the original model. We apply the character expansion to the multiband
superconductor with Josephson couplings in \cref{sec:bexpJ} and discuss it in the light of the
current representation. In \cref{sec:higgs} we present the calculation of the Higgs mass
in terms of current-correlators. We present our conclusions in \cref{sec:conc}.

\section{Standard representation of the model}
\label{sec:model}

We consider a model of $N$ bosonic complex matter fields in three dimensions. The matter
fields are given by
$\psi_\alpha(\bv r)=\left|\psi_\alpha(\bv r)\right|\exp i\theta_\alpha(\bv r)$,
interacting through the electromagnetic vector potential, $\bv A(\bv r)$.
We also allow inter-band Josephson couplings of the matter fields. In the general
case, this is described by a partition function
\begin{equation}
  \mathcal{Z} = \int\mathcal{D}\bv A\left(\prod_\alpha\int\mathcal{D}\psi_\alpha\right) e^{-S},
\end{equation}
where the action is
\begin{align}
  S = \beta\int d^3r\Bigg\{&\frac{1}{2}\sum_\alpha\left(\left|\nabla-ie\bv
A(\bv r)\right)\psi_\alpha(\bv r)\right|^2\nonumber\\&+V\left(\{\left|\psi_\alpha(\bv
r)\right|\}\right)
+\frac{1}{2}{(\nabla\times\bv A(\bv r))}^2 \nonumber\\ -\sum_{\alpha<\beta}\lambda_{\alpha,
\beta}&\left|\psi_\alpha(\bv r)\right|\left|\psi_\beta(\bv r)\right|\cos\left(\theta_\alpha(\bv
r)-\theta_\beta(\bv r)\right)\Bigg\}.
\label{eq:fullHam}
\end{align}
The potential $V$ contains terms that are powers of $\left|\psi_\alpha\right|$.
At this point we employ the phase-only, or London, approximation and choose all
bare stiffnesses, $\left|\psi_\alpha\right|$, equal to unity. Hence, $V$ is
an unimportant constant. We will also focus on equal couplings between all
bands, \textit{i.e.} $\lambda_{\alpha, \beta} = \lambda\;\forall\;\alpha,\beta$.
The action is then given by
\begin{align}
  S = \beta\int d^3r\Bigg\{&\frac{1}{2}\sum_\alpha{\left(\nabla\theta_\alpha(\bv r)-e\bv A(\bv
r)\right)}^2 +\frac{1}{2}{(\nabla\times\bv A(\bv r))}^2 \nonumber\\
&-\lambda\sum_{\alpha<\beta}\cos\left(\theta_\alpha(\bv r)-\theta_\beta(\bv r)\right)\Bigg\}
\end{align}

We regularize this action on a cubic lattice of size $L^3$ by defining the fields
on a discrete set of coordinates $r_\mu\in (1, \ldots, L)$, that is
$\theta_\alpha(\bv r)\rightarrow\theta_{\bv r,
\alpha}$ and $\bv A(\bv r)\rightarrow\bv A_{\bv r}$. On the lattice, the action reads
\begin{align}
  S = \beta\sum_{\bv r}\Bigg\{&-\sum_{\mu, \alpha}\cos\left(\Delta_\mu\theta_{\bv r, \alpha}-eA_{\bv
  r, \mu}\right) +\frac{1}{2}{(\Delta\times\bv A_{\bv r})}^2 \nonumber\\
&-\lambda\sum_{\alpha<\beta}\cos\left(\theta_{\bv r, \alpha}-\theta_{\bv r, \beta}\right)\Bigg\}.
  \label{eq:actionlattice}
\end{align}
Here, we use the cosine function to represent the kinetic term of the continuum Hamiltonian
in a way that preserves the periodic nature of the phases. Alternatively, one may arrive at
\cref{eq:actionlattice} by directly replacing the derivatives in \cref{eq:fullHam} with
the gauge invariant forward difference,
\begin{equation}
  (\nabla-ie\bv A(\bv r))\psi_\alpha(\bv r)\rightarrow\psi_{\bv r+\bvh\mu, \alpha}e^{-ie\bv A_{\bv
  r}}-\psi_{\bv r, \alpha},
\end{equation}
and then taking the London limit as described above.  We discuss the two-dimensional case in
\cref{app:2D}.

In the formulation of \cref{eq:actionlattice} with $\lambda=0$, the model is
known\cite{Smiseth2005,Herland2010} to have one phase transition from a normal state to a
superconducting state in one composite degree of freedom, and $N-1$ phase transitions
from a normal fluid to a superfluid in the remaining degrees of freedom. The reason
for this division into one superconducting and $N-1$ superfluid degrees of freedom
becomes apparent when one correctly identifies the relevant combinations of the phase
fields. The part of the continuum action describing the coupling between the phases
and the gauge field is
\begin{equation}
  S^\prime = \beta\int d\bv r\Bigg\{\frac{1}{2}\sum_\alpha{\left(\nabla\theta_\alpha(\bv r)-e\bv A(\bv
r)\right)}^2\Bigg\}.
\end{equation}
This can be rewritten into\cite{Smiseth2005}
\begin{align}
  S^\prime = \beta\int d\bv r\Bigg\{&\frac{1}{2N}{\left(\sum_\alpha\nabla\theta_\alpha(\bv r)-Ne\bv A(\bv
r)\right)}^2\nonumber\\
&+\frac{1}{2N}\sum_{\alpha<\beta}{\left[\nabla\left(\theta_\alpha-\theta_\beta\right)\right]}^2\Bigg\}.
\label{eq:sumdiffaction}
\end{align}
Hence, the phase combination $\sum_\alpha\theta_\alpha$ will couple to the gauge field, and is
identified as the single charged mode, while all other combinations $\theta_\alpha-\theta_\beta$
do not couple, and are neutral. Note that for $N=1$ only the charged mode remains. Two important
points need to be emphasized. Firstly, the composite variables are not compact in the same sense
that the individual phases are. This means that the composite variables do not support topological
defects by themselves, only composite topological defects. Secondly, the last term in the action of
\cref{eq:sumdiffaction} has $N(N-1)/2$ terms. Therefore, one may not interpret the phase differences
$\theta_\alpha-\theta_\beta$ as independent degrees of freedom. This is because
of the multiple connectedness of the physical space, fluctuations in a single individual
phase induce fluctuations in $N-1$ composite neutral modes, as well as in the charged mode.


In the present form, with $\lambda=0$ and $e$ sufficiently large, this model is known to
have one phase transition in the inverted $3dXY$-universality class, and $N-1$ transitions in the
$3dXY$-universality class at a higher temperature\cite{Smiseth2005,Herland2010}. These transitions
correspond to proliferations of the composite charged mode and the composite neutral modes,
respectively. If the charge is lowered, the charged and neutral transitions will approach each other
in temperature. When they merge, the proliferation of neutral vortices will trigger proliferation of
the charged mode. Consequently, the $N$ phase transitions collapse into a single first-order
transition. This interplay between the charged and neutral sector has been coined a
\textit{preemptive} phase transition \cite{Dahl2008}, and has been verified numerically in
two-component systems in the absence of inter-component Josephson-coupling in several detailed
large-scale Monte Carlo simulations \cite{Kragset2006,Dahl2008,Herland2010}.

In the following Section, we reformulate the model in terms of integer-valued current
fields, considering first the case with zero Josephson-coupling and then move on to include
Josephson coupling. The first case is useful to consider in connecting the results of
previous works mentioned above to the current-formulation.

\section{Current representation of the model}
\label{sec:current_rep}

\subsection{Zero intercomponent Josephson coupling}
\label{sec:bexpJ0}
The basis of the expansion used is a character expansion\cite{Janke1986,Kleinert1989}.
\begin{equation}
  e^{\beta\cos\gamma} = \sum_{b=-\infty}^\infty I_b(\beta)e^{ib\gamma},
  \label{eq:besselid}
\end{equation}
where $I_b(\beta)$ are the modified Bessel functions of integer order. We apply this to the
terms $\exp\beta\cos(\Delta_\mu\theta_{\bv r, \alpha}-e\bv A_{\bv r})$ for each value of
$\bv r$, $\mu$, and $\alpha$. This introduces integer vector fields $\bv b_{\bv r, \alpha}$,
 representing supercurrents. { In fact, the integer vector fields will be the actual physical
 supercurrents of the system\cite{Kleinert1989}.} The low-temperature phase is characterized by a state with
proliferated current-loops on all length scales, while the high-temperature phase only
features small current-loops.

By applying \cref{eq:besselid} to the partition function with \cref{eq:actionlattice} as
the action, and integrating out the phases and the gauge field, details of which may be
found in \cref{app:charexp}, we arrive at the partition function
\begin{align}
  \mathcal{Z} = \sum_{\{\bv b, m\}}&\prod_{\bv r, \alpha}\delta_{\bv\Delta\cdot\bv b_{\bv r, \alpha}, 0}
\prod_{\bv r, \mu, \alpha}I_{b_{\bv r, \alpha, \mu}}(\beta)\nonumber\\
&\prod_{\bv r, \bv r^\prime}e^{-\frac{e^2}{2\beta}\sum_{\alpha, \beta}\bv b_{\bv r, \alpha}\cdot\bv
b_{\bv r^\prime,\beta}D(\bv r - \bv r^\prime)}.
  \label{eq:besselJ0}
\end{align}
This is a model of $N$ current fields, with contact intra-component interactions  parametrized
by the Bessel functions, and long-range intra- and inter-component interactions originating
with  the gauge-field fluctuations, $D(\bv r-\bv r^\prime)$. The constraint
$\bv\Delta\cdot\bv b_{\bv r, \alpha}=0$ forces the currents, $\bv b_{\bv r, \alpha}$
to form closed loops, and implies a non-analytical behavior of each individual component,
and an associated phase transition.

In the current language, the interpretation of the phase transitions explained in the previous
Section is as follows. Consider first a single component model. In the high temperature state, only
the lowest term in the Bessel-function expansion will contribute, and only small loops of
supercurrents  will be present in the system. As the temperature is \textit{lowered} all orders of
the expansion contribute, and the integer currents will proliferate, filling the system with loops
of supercurrent.  In the low temperature state all $b$-fields have proliferated. As temperature is
increased, the proliferated current loops in the charged sector will collapse. Only the neutral
superfluid currents  fill the system, and the state is therefore a metallic
superfluid\cite{Smorgrav2005a}. As temperature is raised further the superfluid currents 
collapse as well, and the system is in the normal metallic state.

\subsection{Non-zero intercomponent Josephson couplings}
\label{sec:bexpJ}

The expansion of \cref{eq:besselid} may also be applied to the Josephson term. { The expansion
is only valid when the argument of the cosine is expanded around zero, the present formulation is
therefore not valid for any ground state which does not fulfill this requirement. In particular, if
the Josephson coupling is negative and sufficiently strong, the phase differences will be locked to
nonzero values \cite{Bojesen2014,Bojesen2015}. For $N=2$ the phases are locked to $\pi$, while for
$N=3$ the ground state of the three phases may form a star-pattern with an accompanying
$\mathrm{Z}_2$ symmetry associated with the two possible chiralities of the
star\cite{Bojesen2014,Bojesen2015}. These cases are not covered by the current-loop formulation
derived from the character-expansion \cref{eq:besselid}. While the above arguments do not constrain
us to only consider all Josephson couplings equal, we may limit our considerations to the case
$\lambda_{\alpha\beta}=\lambda>0$ without loss of generality in the present discussion. Having
universal $\lambda_{\alpha\beta}$ will not allow for any additional physics than simply having
unequal strength of the individual phase lockings, when they are constrained to be all positive.}

Applying the expansion introduces an additional $N(N-1)/2$ integer fields $m_{\bv r, \alpha, \beta}$.
After expanding both the kinetic terms and the Josephson terms, the partition function reads
\begin{align}
  \mathcal{Z} ={}& \int\mathcal{D}\bv A\left(\prod_\alpha\int\mathcal{D}\theta_\alpha\right)\nonumber\\
  \times&\prod_{\bv r, \mu, \alpha}\sum_{b_{\bv r,\mu, \alpha}=-\infty}^\infty I_{b_{\bv r,\mu, \alpha}}(\beta)e^{ib_{\bv r, \mu,
  \alpha}(\Delta_\mu\theta_{\bv r, \alpha}-eA_{\bv r, \bv\mu})}\nonumber\\
  \times&\prod_{\bv r, \alpha<\beta}\sum_{m_{\bv r, \alpha, \beta}=-\infty}^\infty I_{\bv m_{\bv r, \alpha, \beta}}(\beta\lambda)e^{im_{\bv
  r, \alpha, \beta}(\theta_{\bv r, \alpha} - \theta_{\bv r, \beta})}\nonumber\\
  \times&\prod_{\bv r}e^{-\frac{\beta}{2}{(\Delta\times\bv A_{\bv r})}^2}
\end{align}
The effect of the Josephson coupling becomes apparent when we integrate out the phase fields.
The divergences of the $\bv b$-fields will no longer be constrained to zero, but may take any finite
integer value, determined by the value of the $m$-fields. The new constraints read
\begin{equation}
  \bv\Delta\cdot\bv b_{\bv r, \alpha} = \sum_{\beta\neq\alpha}m_{\bv r,
  \alpha, \beta}\;\forall\;\alpha, \bv r,
  \label{eq:Jcons}
\end{equation}
where we have defined $m_{\bv r, \alpha, \beta}=-m_{\bv r, \beta, \alpha}$.
The gauge-term is not coupled directly to the $m$-fields, and we may integrate
it out in the same fashion as before.  The resulting partition function is
\begin{align}
  \mathcal{Z} = \sum_{\{\bv b, m\}}&\prod_{\bv r, \alpha}\delta_{\bv\Delta\cdot\bv b_{\bv r, \alpha},
\sum_{\beta\neq\alpha}m_{\bv r, \alpha, \beta}}\nonumber\\
&\prod_{\bv r, \mu, \alpha}I_{b_{\bv r, \alpha, \mu}}(\beta)  \prod_{\bv r, \alpha<\beta}I_{m_{\bv
r, \alpha, \beta}}(\beta\lambda)\nonumber\\
&\prod_{\bv r, \bv r^\prime}e^{-\frac{e^2}{2\beta}\sum_{\alpha, \beta}\bv b_{\bv r, \alpha}\cdot\bv
b_{\bv r^\prime,\beta}D(\bv r - \bv r^\prime)}
\label{eq:partbJ}
\end{align}

\subsection{Monopoles and phase transitions}

The effect of the $m$-fields is to introduce monopoles into the closed loops of $\bv
b$-currents.  A current of a particular component  (´´color´´) may now terminate at any site. However,
this termination must always be accompanied by a current of another color originating at the same
site. Termination of a current of one component, and the appearance of a current of another component
at the same site represents  an excitation of $\pm 1$ in  $m$. An important
observation is that if one adds the constraints, we have
\begin{equation}
  \sum_\alpha\bv\Delta\cdot\bv b_{\bv r, \alpha} = 0\;\forall\;\bv r.
  \label{eq:sumconstraint}
\end{equation}
This reflects the color changing event stated above, the total current when summing over all
colors is conserved at all sites. It also shows that there is a particular combination of
currents, the sum of all components, which will be divergence-free. The net effect of the
Josephson coupling, pictorially, is to chop up the closed currents of the individual
components and glue them together into closed loops that may change color on any site.

We may expand the partition function first in terms of $m$-fields, and then in terms
of $\lambda$, by using the Bessel-function representation
\begin{equation}
  I_\nu(z) = {\left(\frac{z}{2}\right)}^\nu\sum_{k=0}^\infty
  \frac{{\left(\frac{z}{2}\right)}^{2k}}{k!(\nu+k)!}
\end{equation}
This demonstrates that the partition function consists of a single term with zero divergence
on all sites, which we know has one or more phase transitions from a superconducting
superfluid state into a non-superconducting normal fluid, and many terms where the divergence of
$\bv b_{\bv r,\alpha}$ is finite on any number of sites.

Let us now consider two limits, and assume $e$ is large, so that there is no preemptive effect for
$\lambda=0$. For $\lambda=0$, it is evident that only $m=0$ will
contribute, and we are left with only divergenceless terms, and hence the behaviour described
previously. The other limit is $\lambda\rightarrow\infty$. In this case we must examine the
asymptotic form of the Bessel functions, which to leading order in the argument is
\begin{equation}
  I_m(z) \sim \frac{e^z}{\sqrt{2\pi z}},
\end{equation}
\textit{i.e.} independent of $m$, and the monopole field will fluctuate strongly,
causing the zero-divergence constraint on each component to be removed. The only remaining
constraint on the current fields pertains to the composite  current
$\sum_\alpha\bv b_\alpha$, which is divergence-free. The interpretation
of this is that the phase transitions in the  $N-1$
superfluid modes are converted to crossovers by the Josephson coupling, while the single
superconducting mode still undergoes a genuine phase transition. The neutral crossover
will be far removed from the charged phase transition in this limit, and the remaining
fluctuations in the neutral sector will be almost completely suppressed. There is no
possibility of any interference between the sectors, and therefore no preemptive phase
transition. The phase transition in the charged sector will therefore be in the universality
class of the inverted $3DXY$ phase transition.

For intermediate and small values of $\lambda$, the effect of the Josephson coupling on
the interplay between the charged and neutral sectors is quite subtle in the present
formulation, and will be discussed in the following section.

\subsection{Charged and neutral currents}

We start with the action where the phase
sum and phase differences have been separated, \cref{eq:sumdiffaction}, with a
Josephson coupling included. To simplify the notation, we introduce composite fields
$\Theta\equiv\sum_\alpha\theta_\alpha$ and
$\vartheta_{\alpha\beta}\equiv\theta_\alpha-\theta_\beta$.
The lattice action then reads
\begin{align}
  S = \beta\sum_{\bv r}\Bigg\{&-\sum_\mu\cos\left(\Delta_\mu\Theta_{\bv r}-NeA_{\bv r, \mu}\right)\nonumber\\
                         &-\sum_{\mu, \alpha<\beta}\cos\left(\Delta_\mu\vartheta_{\bv r,
                         \alpha\beta}\right)-\lambda\sum_{\alpha<\beta}\cos\left(\vartheta_{\bv r,
                       \alpha\beta}\right)\nonumber\\&+\frac{1}{2}{\left(\bv\Delta\times\bv A_{\bv r}\right)}^2\Bigg\}.
\label{eq:actioncompositelattice}
\end{align}
One may arrive at this form by defining the composite fields in \cref{eq:sumdiffaction}, then
use the Villain approximation on the original action of \cref{eq:actionlattice}, rewrite the resulting action into one with the composite fields, then reverse the Villain approximation.

In \cref{eq:actioncompositelattice}, there is one charged mode and $N(N-1)/2$ neutral 
modes, while the original theory has $N$ degrees of freedom. There is therefore an 
excess of $(N-1)(N-2)/2$ degrees of freedom. (Note that there are no redundant modes 
for $N=1$ and $N=2$). Therefore, not all of the phase differences are independent when 
$N>2$. Consider the case $N=3$, where one may form the phase differences
$\theta_1-\theta_2$, $\theta_2-\theta_3$ and $\theta_1-\theta_3$, but
$\theta_1-\theta_3 = (\theta_1-\theta_2)+(\theta_2-\theta_3)$. It suffices
to include the phase differences $\vartheta_{12}$ and $\vartheta_{23}$.

This may be generalized to arbitrary $N$. Identify all $\theta_{\alpha\beta}$ where
\begin{equation}
\{(\alpha,\beta)|\alpha\in(1,\ldots,N-1)\land\beta=\alpha+1\}.
  \label{eq:realdof}
\end{equation}
Then, all $\theta_{\alpha\beta}$ where
\begin{equation}
\{(\alpha,\beta)|\alpha\in(1,\ldots, N-2)\land\beta\in(\alpha+2, \ldots,
N)\}
  \label{eq:redundantdof}
\end{equation}
may be constructed by adding up the intermediate phase differences, that is
$\vartheta_{\alpha\beta}=\vartheta_{\alpha, \alpha+1} + \vartheta_{\alpha+1, \alpha+2} + \cdots
\vartheta_{\beta-1, \beta}$. With this in mind, we may write out the partition function in terms of
the charged and neutral modes
\begin{align}
  \mathcal{Z} =
  &\int\mathcal{D}\Theta\left(\prod_{\alpha<\beta}\int\mathcal{D}\vartheta_{\alpha\beta}\right)\nonumber\\
  &\times\left(\prod_{\alpha=1}^{N-1}\prod_{\beta=\alpha+2}^N\delta\left(\vartheta_{\alpha\beta}-\sum_{\eta=\alpha}^{\beta-1}\vartheta_{\eta,\eta+1}\right)\right)e^{S}
  \label{eq:partredundant}
\end{align}
where $S$ is the action of \cref{eq:actioncompositelattice}.

As an illustration,  we perform the character expansion on the model where the charged and neutral sectors have been separated, for the special cases $N=2$ and $N=3$.

For $N=2$ there are no redundant variables, and we have the two composite variables
$\Theta\equiv\theta_1+\theta_2$ and $\vartheta\equiv\theta_1-\theta_2$. Using the identity
\cref{eq:besselid}, and integrating out the phases and gauge field, we obtain
\begin{align}
  \mathcal{Z} = \sum_{\{\bv B, \boldsymbol{\mathcal{B}}, m\}}&\prod_{\bv r}\delta_{\bv\Delta\cdot\bv B_{\bv r},0}
  \delta_{\bv\Delta\cdot\boldsymbol{\mathcal{B}}_{\bv r},m_{\bv r}}\nonumber\\
  &\prod_{\bv r, \mu}I_{B_{\bv r, \mu}}(\beta)I_{\mathcal{B}_{\bv r, \mu}}(\beta)\prod_{\bv r}I_{m_{\bv r}}(\beta\lambda)\nonumber\\
  \prod_{\bv r, \bv r^\prime}\exp&\left\{-\frac{({Ne)}^2}{2\beta}\bv B_{\bv r}\cdot\bv B_{\bv r^\prime}D(\bv r - \bv r^\prime)\right\}.
	\label{Z_2}
\end{align}
Here, $\bv B$ is the charged current field associated with $\Theta$, while
$\boldsymbol{\mathcal{B}}$ is the neutral current field associated with $\vartheta$.

In this formulation, it is immediately clear that the  model features two integer
vector-field degrees of freedom,
one which has long-range interactions mediated by the gauge field, and one with  contact
interactions. The neutral current field has its constraint removed by the $m$-field, while the
charged field is still constrained to be divergenceless. Hence, the model will feature a single
phase transition in the charged sector driven by the collapse of closed loops of charged
currents, while the transition of the neutral sector is converted to a crossover
by the complete removal of constraints on $\boldsymbol{\mathcal{B}}$.

Let us consider this in a bit more detail. In \cref{Z_2}, we may perform the summation
over the fields $m \in \mathbb{Z}$. Since we have that
$\bv\Delta\cdot\boldsymbol{\mathcal{B}}_{\bv r}\in \mathbb{Z}$ as
well, the summation over the $m$'s will guarantee that the constraint is satisfied
for some value of $m$, such that the summation over $m$ effectively removes the constraints
on $\bv\Delta\cdot\boldsymbol{\mathcal{B}}_{\bv r} $.  Hence, we have $\sum_{\{m\}}
\delta_{\bv\Delta\cdot\boldsymbol{\mathcal{B}}_{\bv r},m_{\bv r}} \prod_{\bv r}I_{m_{\bv
r}}(\beta\lambda) = \prod_{\bv r} I_{\bv\Delta\cdot\boldsymbol{\mathcal{B}}_{\bv r}
}(\beta\lambda)$, with no constraints on $\bv\Delta\cdot\boldsymbol{\mathcal{B}}_{\bv r}$.
We may thus perform the now unconstrained summation of the field
$\boldsymbol{\mathcal{B}}_{\bv r}$, namely
\begin{equation}
\sum_{\{ \boldsymbol{\mathcal{B}}\}}
\left( \prod_{\bv r, \mu} I_{\mathcal{B}_{\bv r, \mu}}(\beta) \right)
\left(\prod_{\bv r} I_{\bv\Delta\cdot\boldsymbol{\mathcal{B}}_{\bv r}  }(\beta\lambda) \right)
= F(\beta,\lambda),
\label{eq:neutralanalytic}
\end{equation}
where $F$ is an analytic function of its arguments. This may be seen by mapping the left hand
side of \cref{eq:neutralanalytic} to a Villain model,  using the approximation\cite{Janke1986}
\begin{equation}
  \frac{I_b(x)}{I_0(x)}\approx \frac{1}{\left|b\right|!}e^{\log(\beta/2)\left|b\right|}.
\end{equation}
This  may be rewritten as a gaussian provided $\beta$ is sufficiently small so that
contributions $\left|b\right|>1$ are small,
\begin{equation}
  \frac{I_b(x)}{I_0(x)}\approx e^{\frac{-b^2}{2\beta^\prime}},
\end{equation}
where $\beta^\prime$ is a renormalized coupling constant, and we find
\begin{equation}
F(\beta,\lambda)=
\sum_{\{ \boldsymbol{\mathcal{B}}\}}
  \left( \prod_{\bv r, \mu}\exp{\frac{-\mathcal{B}_{\bv r, \mu}^2}{2\beta^\prime}}\right)
  \left(\prod_{\bv r} \exp{\frac{-\left(\bv\Delta\cdot\boldsymbol{\mathcal{B}}_{\bv r}\right)^2}{2\lambda\beta^\prime}}\right),
  \label{eq:neutralanalyticgauss}.
\end{equation}
Since there are no constraints  $\boldsymbol{\mathcal{B}}$,
this demonstrates that \cref{eq:neutralanalytic}  essentially is a
discrete Gaussian theory, and the neutral sector therefore does not suffer any phase
transition. This point may be further corroborated by going back to the formulation of
Eq. \ref{eq:actioncompositelattice}.  The neutral sector of the action
is seen to be identical to that of an $XY$ spin-model in an external magnetic field, with field
strength $\lambda$. Any $\lambda \neq 0$ converts the phase transition, from a low-temperature
ferromagnetic state to a high-temperature paramagnetic state, into to a crossover from an ordered to
a disordered system. Note also that in the limit $\lambda=0$, the Bessel function will revert to
$I_{\bv\Delta\cdot\boldsymbol{\mathcal{B}}}(0)=\delta_{\bv\Delta\cdot\boldsymbol{\mathcal{B}}, 0}$,
and the non-analytical constraint is re-introduced.

{We emphasize that although the above argument utilized a Villain-approximation to the Bessel-functions, the conclusion that the phase-transition is wiped out in the neutral sector by introducing monopoles (Josephson-coupling) does not depend on this approximation. At any rate, 
a Villain-approximation to the XY-model does not change the symmetry of the problem or the 
character of phase transitions. What is crucial is the introduction of monopoles and the 
ensuing removal of constraints on the neutral currents.}  
 

The total partition function for the entire system is thus given by
 \begin{eqnarray}
  \mathcal{Z} & = &  F(\beta, \lambda)
	\sum_{\{\bv B \}} \prod_{\bv r}\delta_{\bv\Delta\cdot\bv B_{\bv r},0} \prod_{\bv r, \mu}I_{B_{\bv r, \mu}}(\beta)  \nonumber \\
	& & \prod_{\bv r, \bv r^\prime}\exp \left\{-\frac{({Ne)}^2}{2\beta}\bv B_{\bv r}\cdot\bv B_{\bv r^\prime}D(\bv r - \bv r^\prime)\right\}.
	\label{Z_3}
\end{eqnarray}
The phase-transition in the neutral sector is converted to a crossover, and there are no longer any
\textit{critical} fluctuations associated with disordering the neutral sector, unlike the case
$\lambda=0$. This occurs as soon as $\lambda$ is finite, however small. However, even without a phase
transition and associated critical fluctuations, there will still be a crossover with associated
fluctuations in its vicinity. Hence, the preemptive first-order phase transition in the charged sector, which occurs for $\lambda=0$, may still take place provided  $\lambda$ sufficiently small.

The argument  is as follows. In the preemptive scenario for $\lambda=0$, fluctuations in the neutral
and charged sectors increase as $T$ is increased from below in the fully ordered state. The charged
sector influences the fluctuations in the neutral sector and vice versa, such that the putative
continuous transitions in these sectors are preempted by a common
first order phase transition \cite{Smiseth2005,Herland2010}. The
important point to realize is that neither of the sectors actually reach criticality,
since there are no critical fluctuations at the preemptive first-order phase transition.

We may have the same scenario occurring with finite but small $\lambda$.
A necessary requirement is that the gauge-charge $e$ is not too large, such that
gauge-field fluctuations are not so large as to separate the phase-transitions
in the charged and the neutral sector too much \cite{Smiseth2005,Herland2010}.
The key point is that the inclusion of Josephson-couplings converts the phase
transition in the neutral sector to a crossover in exactly the same way that
the ferromagnetic-paramagnetic phase transition in the $3DXY$ model is converted
to a crossover by the inclusion of a magnetic field coupling linearly to the
$XY$-spins, cf. Eq \ref{eq:actioncompositelattice}.
This leaves only a phase-transition in the charged sector, but it does not completely
suppress fluctuations in the neutral sector. It merely cuts the fluctuations off on
a length-scale given by the Josephson-length $1/\lambda$, thereby preventing them
from becoming critical. As temperature is increased, the neutral sector approaches
its crossover region,
with increasingly large fluctuations. At the same time, the charged sector approaches
its putative inverted-$3dXY$ fixed point. Provided that the crossover region of the
neutral sector and the fixed point of the charged sector are sufficiently close, the
fluctuations in both sectors may still strongly influence each other, and a first-order
preemptive phase transition may still occur in the charged sector. This is consistent
with  recent numerical work\cite{Sellin2016}, which observed a first order phase transition
in multi-band superconductors with weak Josephson-coupling in Monte-Carlo simulations
using the original $\mathrm{U}(1)$ phases.

For $N=3$ we must consider carefully the redundant variable, $\vartheta_{13} =
\vartheta_{12}+\vartheta_{23}$. The partition function, prior to integration of
the phases and the gauge field reads
\begin{align}
  \mathcal{Z} ={}&
  \int\mathcal{D}\Theta\left(\prod_{\alpha<\beta}\int\mathcal{D}\vartheta_{\alpha\beta}\right)\delta(\vartheta_{13}-\vartheta_{12}-\vartheta_{23})\nonumber\\
  \times&\prod_{\bv r, \mu}\sum_{B_{\bv r,\mu}=-\infty}^\infty I_{B_{\bv r,\mu}}(\beta)e^{iB_{\bv r,
  \mu}(\Delta_\mu\Theta_{\bv r}-NeA_{\bv r, \bv\mu})}\nonumber\\
  \times&\prod_{\substack{\bv r, \mu\\\alpha<\beta}}\sum_{\mathcal{B}_{\bv r, \mu,
  \alpha\beta}=-\infty}^\infty I_{\mathcal{B}_{\bv r,\mu, \alpha\beta}}(\beta)e^{i\mathcal{B}_{\bv
  r, \mu, \alpha\beta}\Delta_\mu\vartheta_{\bv r, \alpha\beta}}\nonumber\\
  \times&\prod_{\bv r, \alpha<\beta}\sum_{m_{\bv r, \alpha\beta}=-\infty}^\infty I_{\bv m_{\bv r, \alpha, \beta}}(\beta\lambda)e^{im_{\bv
r, \alpha, \beta}\vartheta_{\alpha\beta}}\nonumber\\
  \times&\prod_{\bv r}e^{-\frac{\beta}{2}{(\Delta\times\bv A_{\bv r})}^2}.
\end{align}
Again, $\bv B$ is the charged current associated with $\Theta$, while
$\boldsymbol{\mathcal{B}}_{\alpha\beta}$ are the neutral currents associated with
$\vartheta_{\alpha\beta}$. The $\delta$-function is included to account for the redundancy of the
composite phase representation.

We now proceed with the integration of phases and gauge field, taking care to integrate
out the redundant phase first. The partition function may then be written as
\begin{align}
  \mathcal{Z} = \sum_{\{\bv B, \boldsymbol{\mathcal{B}}, m\}}&\prod_{\bv r}\delta_{\bv\Delta\cdot\bv B_{\bv r},0}\prod_{\bv r, \mu}I_{B_{\bv r, \mu}}(\beta)\nonumber\\
  &\prod_{\bv r}\delta_{\bv\Delta\cdot\boldsymbol{\mathcal{B}}_{\bv r, 12}+\bv\Delta\cdot\boldsymbol{\mathcal{B}}_{\bv r, 13},m_{\bv r, 12}+m_{\bv r, 13}}\nonumber\\
  &\prod_{\bv r}\delta_{\bv\Delta\cdot\boldsymbol{\mathcal{B}}_{\bv r, 23}+\bv\Delta\cdot\boldsymbol{\mathcal{B}}_{\bv r, 13},m_{\bv r, 23}+m_{\bv r, 13}}\nonumber\\
  &\prod_{\substack{\bv r, \mu\\\alpha<\beta}}I_{\boldsymbol{\mathcal{B}}_{\bv r, \mu, \alpha\beta}}(\beta)
  \prod_{\substack{\bv r\\\alpha<\beta}}I_{m_{\bv r, \alpha\beta}}(\beta\lambda)\nonumber\\
  \prod_{\bv r, \bv r^\prime}\exp&\left\{-\frac{({Ne)}^2}{2\beta}\bv B_{\bv r}\cdot\bv B_{\bv r^\prime}D(\bv r - \bv r^\prime)\right\}.
  \label{eq:partbJsep}
\end{align}
This is a model of a single gauge coupled supercurrent $\bv B$ which are constrained to form closed
loops, and three superfluid currents
$\boldsymbol{\mathcal{B}}_{12}$, $\boldsymbol{\mathcal{B}}_{23}$ and
$\boldsymbol{\mathcal{B}}_{13}$ which are not constrained to form closed loops.
The three superfluid currents are not independent, as is seen from the two constraints on them.
As in the case $N=2$, the summation over the $m$-fields may be performed, eliminating the
constraints on the fields $\boldsymbol{\mathcal{B}}_{\bv r, \mu, \alpha\beta}$, after which the
\textit{unconstrained} summation over these fields may be performed. As for $N=2$, this yields
multiplicative analytic factors in the partition function, and the phase transitions in the neutral
sectors will be converted to crossovers. Given that the crossovers in the neutral sectors and
the charged fixed point have sufficient overlap, the system may still feature a
single preemptive first-order phase transition arising from the interplay between the charged and
neutral modes. Furthermore, the inclusion of the additional degree of freedom  enhances the
combined fluctuations of the neutral mode at a given Josephson coupling, $\lambda$, and therefore
strengthens the preemptive first-order transition. This is  consistent with the results of
recent numerical work\cite{Sellin2016}.


\subsection{Preemptive effect and current-loop interactions}

\begin{figure}
\includegraphics[width=\columnwidth]{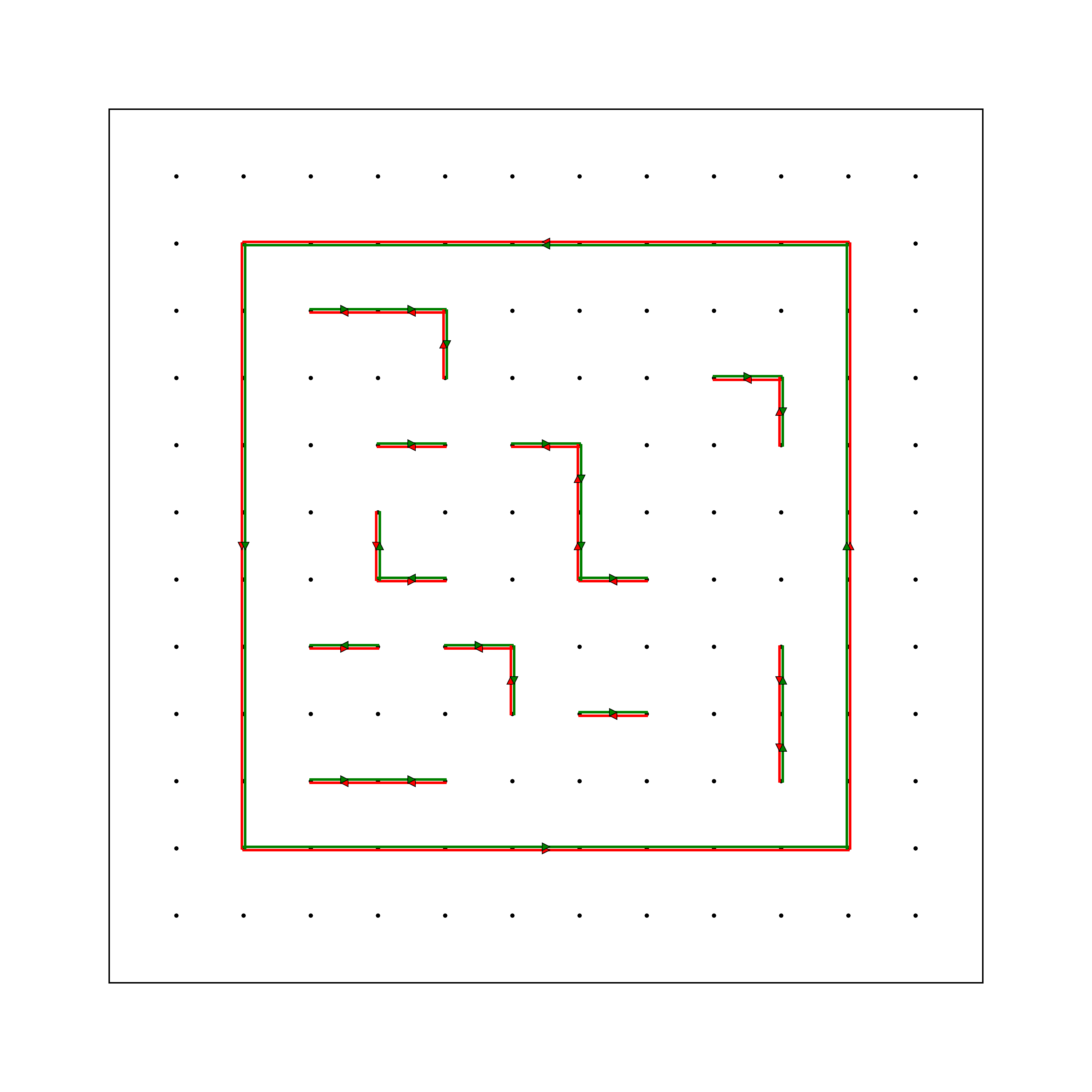}
\caption{Example of a current-loop configuration in the case of strong inter-band Josephson
  coupling, when there is no screening of the charged-current interaction. Red and green lines
  represent currents of the individual fields $\bv b_i$ flowing in the direction indicated by the
  arrows. Charged and neutral currents are therefore represented by overlapping red and green lines
  flowing either in the same or the opposite direction, respectively. The configuration shown
  represents a snapshot close to the charged transition, where a closed loop of charged current
  encircles pieces of a tightly bound composite neutral current. As there is no interaction between pieces of
  charged and neutral current, the inverted-$3DXY$ transition of the charged sector is not
  influenced by the tightly bound composite neutral currents.}
\label{fig:unscreened}
\end{figure}

\begin{figure}
\includegraphics[width=\columnwidth]{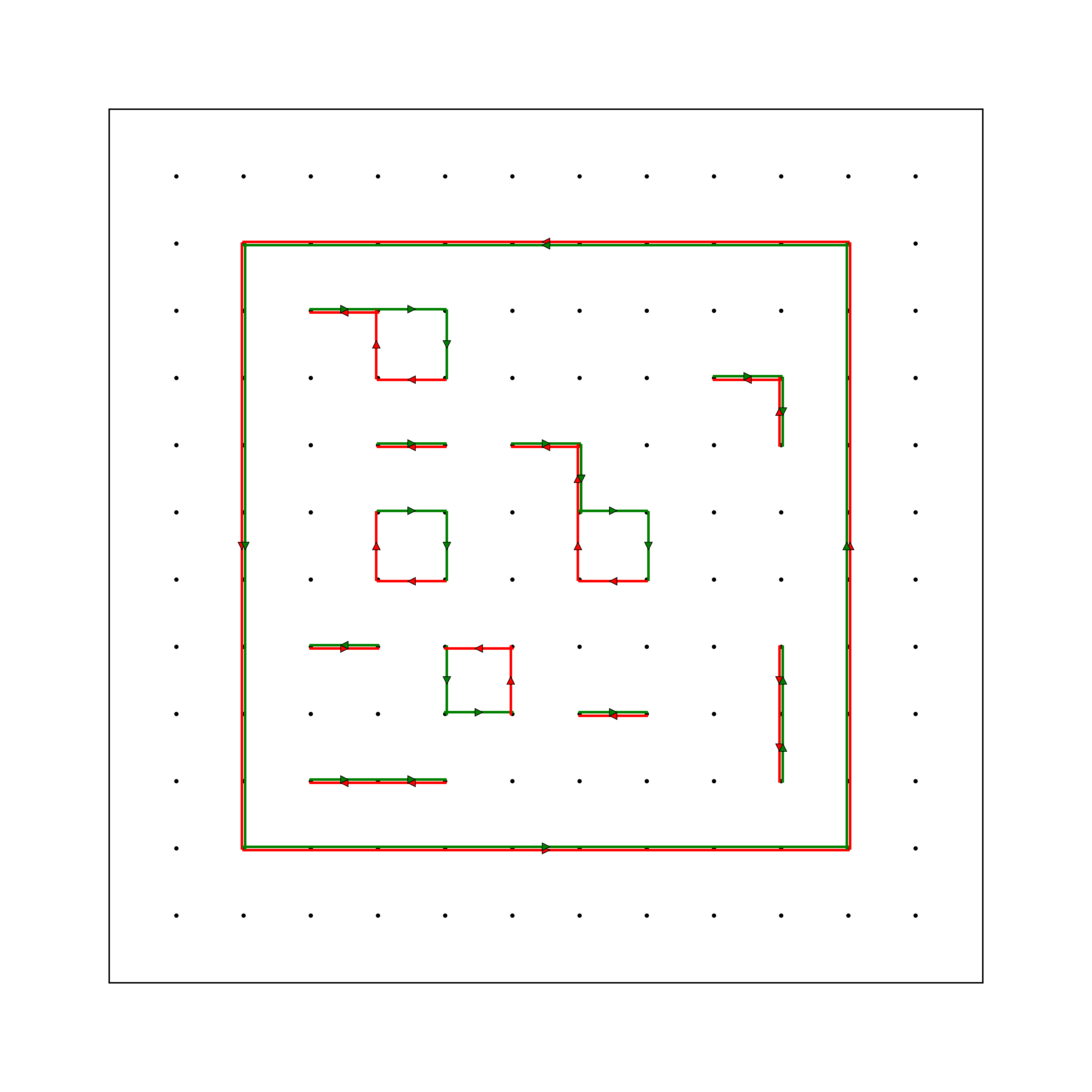}
\caption{Example of a current-loop configuration in the case of weak, but non-zero, inter-band Josephson
  coupling. Red and green lines represent currents of the individual fields $\bv b_i$ flowing in the direction indicated by the
  arrows. Charged and neutral currents are therefore represented by overlapping red and green lines
  flowing either in the same or the opposite direction, respectively. The present configuration show
  the same loop of charged current ecircling pieces of neutral current, as shown in
  \cref{fig:unscreened}. However, in the case of weak inter-band Josephson coupling, the individual
  currents will fluctuate away from the neutral-current configuration slightly close to the neutral
  crossover. { The screening of the interaction between the segments of the outer charged composite current is accounted for by removing all tightly bound counterflowing currents in the interior,
	leaving only closed loops that change color an even number of times as the loops are traversed. These closed loops screen the charged-current interaction and
  may effectively change the sign of the interaction between the segments of charged currents,
	as explained in the text. }This in turn may cause the transition of the charged sector to
  turn first order. 
  }
\label{fig:screened}
\end{figure}

{In this subsection, we discuss further the preemptive scenario discussed above, interpreting
it in terms of renormalizations of current-current interactions. This provides a dual picture
to the physical picture of the first-order phase transition presented in   
Ref. \onlinecite{Sellin2016}.  

The preemptive phase transition
may be understood in the current-loop picture by considering the effect of the monopoles on the
neutral counter-flowing current sector (facilitated by the presence of monopoles, i.e. Josephson coupling), and how this in turn influences the interaction between the charged co-flowing currents
which interact via the fluctuating gauge-field.

Consider first the current-loop excitations allowed by \cref{eq:partbJ} for the 
case $N=2$. The lowest order configurations in the individual fields are closed loops of a 
single color. On top of these one may add monopoles, such that one has closed loops that 
change color twice before completing a closed loop. The presence of the Josephson coupling 
also allows for small dumbbells of counter-flowing currents with a monopole at one end and 
an anti-monopole at the other end. The gauge field will bind loops of co-flowing currents 
together, creating small loops of both colors flowing in the same direction. At high 
temperatures, the co-flowing currents only form small closed loops, and the system is
non-superconducting. Barring any influence from the neutral sector, they will proliferate 
in an inverted $3DXY$-transition \cite{PhysRevLett.47.1556} at some critical temperature. 
If the charge, or the Josephson coupling, is sufficiently strong, there will be no 
significant fluctuations in the neutral sector that may influence this. The co-flowing 
current loops simply proliferate in a background of \textit{only} tightly bound 
counter-flowing currents, with which they do not interact at all. The only way they 
can interact is if a counter-flowing composite current locally dissociates into individual
currents on length scales below the Josephson length, which needs to be large enough. This
will not happen if either the Josephson coupling is sufficiently strong, or if the charge is
sufficiently large so that the charged transition is separated sufficiently from the neutral
crossover.

\cref{fig:unscreened,fig:screened} show simple representations of current configurations as the
transition occurs in the two scenarios. For simplicity the illustration is given in two spatial dimensions. In \cref{fig:unscreened}, we show the case of having a
sufficiently strong Josephson coupling. A generic snapshot of a single loop of charged current is
shown, represented by two co-flowing red and green lines, surrounding a gas of tightly bound pieces
of counter-flowing neutral currents. As there are no individual red or green lines, there will be no
interactions between the loop of composite charged current and the small pieces of 
composite neutral current, and hence
no renormalization of the interactions in the charged sector. The loops of charged current will
therefore proliferate in an inverted $3DXY$-transition \cite{PhysRevLett.47.1556} as the temperature is lowered. In
\cref{fig:screened}, the situation is different. Here, the Josephson coupling is sufficiently low,
or alternatively the Josephson length is sufficiently large, so that the individual pieces of
current may undergo local dissociations of the tighly bound counter-flowing configurations.
These individual pieces of currents, represented by only red or green lines, will interact with 
the loop of charged current, and may therefore influence the  proliferation of composite charged
current loops.

The current loops are dual objects to vortex loops. It is known that there is a precise
correspondence between the sign of vortex interactions and the character of the
phase transition in superconductors. Namely, attractive interactions between vortices
leads to a first-order phase transition, while repulsive vortex interactions
lead to second order phase transitions \cite{Mo2002,Sellin2016}. Therefore, an alternative 
natural way of interpreting the preemptive first-order phase-transition in the dual picture,
is that neutral counter-flowing currents on the co-flowing charged currents 
screen or overscreen the interactions between the latter, effectively changing 
the sign of the interactions between charged current-segments. 

{ With reference to \cref{fig:screened}, we elaborate briefly on how the configurations depicted 
there may cause attractive interactions between composite charged current-segments. Note that 
the screening is accounted for entirely by removing all tightly bound counter-flowing currents,
leaving only the closed color-changing loops. The relevant screening fluctuations are therefore
complicated collective phase-fluctuations amounting to inserting closed color-changing loops
in the problem. Loops which interact attractively with the composite charged current-segments
will have a larger Boltzmann-weight in the dual action than those that attract repulsively,
and they will therefore dominate the configurations where many closed current-changing 
(originating with tightly bound counter-flowing currents) are present. This attraction may 
cause an effective attraction between the charged composite current segments, via the 
attraction to the closed current-changing loops. An identical physical picture holds when 
working with the dual objects to the currents, namely vortices.}    

To summarize, the basic mechanism causing a first-order phase transition is the influence of 
partial decomposition of composite neutral
currents on the interaction between charged composite currents, equivalently the influence of partial decomposition of composite
neutral vortices on the interaction between composite charged vortices. These pictures are
particular dual manifestations of the general concept of a preemptive first order phase transition.
In such a transition, a putative second order phase transition associated with proliferation of
topological defects in a given order parameter, is converted to a first order phase transition
preemptively  by strong fluctuations (not necessarily critical) in some other field.  }

\section{Current correlations and the Higgs mechanism}
\label{sec:higgs}
The defining characteristic of the inverted 3D$XY$-transition in the charged sector
is a spontaneous $U(1)$ gauge-symmetry breaking associated with the gauge field $\bv A$
becoming massive as the system crosses the transition point of the metallic state into the
superconducting state. In this section, we investigate how the onset of the mass $m_{\bv A}$
of the photon (the Higgs mass),  which is equivalent to the Meissner effect of the
superconductor, comes about as result of a non-analytic change in the infrared
properties of the current-correlations of the system. $m_{\bv A}$ is found from the
limiting form of the gauge-field correlation function
\begin{eqnarray}
\langle A_{\bv q}^\mu A_{-\bv q}^ \nu\rangle \sim
\frac{1}{q^2+m_{\bv A}^2}.
\end{eqnarray}
To calculate $\langle A_{\bv q}^\mu A_{-\bv q}^ \nu\rangle $, we consider the action
of the charged sector given on the form \cref{eq:Sq} before
integrating out the gauge field, and insert source terms
source $\bv J_{\bv q}$,
 \begin{align}
   S_J = \sum_{\bv q}\Bigg[&\frac{ie}{2}\sum_\alpha\bv b_{\bv q, \alpha}\cdot\bv A_{\bv
 -q}+\frac{ie}{2}\sum_\alpha\bv b_{\bv -q, \alpha}\cdot\bv A_{\bv q}\nonumber\\
 &+\frac{\beta}{2}\left|\bv Q_{\bv q}\right|^2\bv A_{\bv q}\cdot\bv A_{-\bv q}\nonumber\\
 &+\frac{1}{2}\left(\bv J_{\bv q}\cdot\bv A_{-\bv q}+\bv J_{-\bv q}\cdot\bv A_{\bv q}\right)\Bigg].
\end{align}
which in turn may be written on form
\begin{align}
  S_J = \sum_{\bv q}\Bigg[&\Big(\bv A_{\bv q} + \frac{1}{2}\left(\bv J_{\bv q} + ie\sum_\alpha\bv
b_{\bv q, \alpha}\right)D_{\bv
      q}^{-1}\Big)D_{\bv q}\nonumber\\
      &\times\Big(\bv A_{-\bv q} + \frac{1}{2}\left(\bv J_{-\bv q}+ie\sum_\alpha\bv b_{-\bv q,
  \alpha}\right)D_{\bv
      q}^{-1}\Big)\nonumber\\
      &+ -\frac{1}{4}\Big(\bv J_{\bv q}+ie\sum_\alpha\bv b_{\bv q, \alpha}\Big)D_{\bv
    q}^{-1}\nonumber\\
  &\times\Big(\bv J_{-\bv q}+ie\sum_\beta\bv b_{-\bv q, \beta}\Big)\Bigg].
\end{align}
Here, $D_{\bv q} = \beta\left|\bv Q_{\bv q}\right|^2/2$ as before. After shifting and
integrating the gauge field, we have
\begin{align}
  S_J = -\sum_{\bv q}\Bigg[&\frac{1}{2\beta\left|\bv Q_{\bv q}\right|^2}\Big(J_{\bv q}^\mu
  P_T^{\mu\nu}J_{-\bv q}^\nu-e^2\sum_{\alpha\beta} b_{\bv q, \alpha}^\mu b_{-\bv q}^\mu \nonumber\\
  &+ie\sum_\alpha\left(J_{-\bv q}^\mu b_{\bv q, \alpha}^\mu + J_{\bv q}^\mu b_{-\bv q, \alpha}^\mu\right)\Big)\Bigg],
\end{align}
where repeated indices are summed over, and $P_T^{\mu\nu}$ is the
transverse projection operator
\begin{equation}
  P_T^{\mu\nu} = \delta^{\mu\nu} - \frac{Q^\mu_{\bv q}Q^\nu_{-\bv q}}{\left|\bv Q_{\bv q}\right|^2}
\end{equation}

The gauge-field correlator is then given by
\begin{widetext}
\begin{align}
  \langle A_{\bv q}^\mu A_{-\bv q}^ \nu\rangle ={}&
  \frac{1}{\mathcal{Z}_0}\frac{\delta^2\mathcal{Z}_J}{\delta
  J_{-\bv q,\mu}\delta J_{\bv q,\nu}}\Bigg|_{\bv J=0}\nonumber\\
  ={}& \frac{1}{\mathcal{Z}_0}\sum_{\{\bv b, m\}}\prod_{\bv r, \alpha}\delta_{\bv\Delta\cdot\bv b_{\bv r, \alpha},
\sum_{\beta\neq\alpha}\epsilon_{\alpha\beta}m_{\bv r, \alpha, \beta}}
\prod_{\bv r, \mu, \alpha}I_{b_{\bv r, \alpha, \mu}}(\beta)  \prod_{\bv r, \alpha<\beta}I_{m_{\bv
r, \alpha, \beta}}(\beta\lambda)\nonumber\\
&\times\left(-\frac{\delta^2S_J}{\delta\bv J_{-\bv q}^\mu\delta\bv J_{\bv q}^\nu}
-\frac{\delta S_J}{\delta\bv J_{-\bv q}^\mu}\frac{\delta S_J}{\delta\bv J_{\bv q}^\nu}
\right)e^{-S_J}\Bigg|_{\bv J = 0}
\label{eq:Aqderiv}
\end{align}
\end{widetext}
Here, $\mathcal{Z}_0$ is the partition function with the sources set to zero.
The functional derivatives of the action is given by
\begin{equation}
  -\frac{\delta S_J}{\delta\bv J_{\bv q}^\nu} = \frac{1}{\beta\left|\bv Q_{\bv q}\right|^2}
  (J_{-\bv q}^\nu P_T^{\mu\nu} + ie\sum_\alpha b_{-\bv q, \alpha}^\nu)
\end{equation}
and
\begin{equation}
  -\frac{\delta^2S_J}{\delta\bv J_{-\bv q}^\mu\delta\bv J_{\bv q}^\nu} = \frac{1}{\beta\left|\bv
  Q_{\bv q}\right|^2}P_T^{\mu\nu}.
\end{equation}
Inserting this into \cref{eq:Aqderiv} and setting the currents to zero, we have
\begin{equation}
  \langle A_{\bv q}^\mu A_{-\bv q}^ \nu\rangle = \frac{P_T^{\mu\nu}}{\beta\left|\bv Q_{\bv
  q}\right|^2}-\frac{e^2}{\beta^2\left|\bv Q_{\bv q}\right|^4}\langle\sum_{\alpha\beta}b_{\bv q,
  \alpha}^\mu b_{-\bv q, \beta}^\nu\rangle
\end{equation}
Setting $\nu=\mu$ and summing over $\mu$ yields the relevant correlator
\begin{equation}
  \langle \bv A_{\bv q}\cdot \bv A_{-\bv q}\rangle = \frac{1}{\beta\left|\bv Q_{\bv
  q}\right|^2}\left(2-\frac{e^2}{\beta\left|\bv Q_{\bv q}\right|^2}\langle\bv{B}_{\bv
    q}\cdot\bv{B}_{-\bv q}\rangle\right),
  \label{eq:Acorrapp}
\end{equation}
where we have defined $\langle \bv{B}_{\bv q}\cdot\bv{B}_{-\bv q}\rangle=\langle\sum_{\alpha\beta}\bv b_{\bv q,
  \alpha}\cdot\bv b_{-\bv q, \beta}\rangle$

The effective gauge field mass is given by the zero momentum limit of the inverse propagator,
\begin{eqnarray}
m_{\bv A}^2
& = & \lim_{\bv q\rightarrow 0}\frac{2}{\beta\langle\bv A_{\bv q}\bv A_{-\bv q}\rangle}
\end{eqnarray}
As is seen from \cref{eq:Acorrapp}, the relevant combination of current-field correlators
is the superconducting current, while charge-neutral currents do not appear in the
expression. The current-correlator may be interpreted as the helicity modulus,
which at a charged fixed point has a non-analytic behavior of the term proportional
to $q^2$. We expect the leading behavior to be\cite{Hove2000}
\begin{equation}
  \lim_{\bv q\rightarrow 0}\frac{e^2}{2\beta}\langle \bv{B}_{\bv q}\cdot\bv{B}_{-\bv q}\rangle\sim\left\{
    \begin{array}{@{}ll@{}}
      (1- C_2(T))q^2,&   T>T_C. \\
      q^2 - C_3(T)q^{2+\eta_{\bv A}},&   T=T_C. \\
      q^2 - C_4(T)q^4,&   T<T_C. \\
    \end{array}\right.
\end{equation}
The result given above is dual to an expression for the gauge-mass in terms
of correlation function of topological defects of the superconducting order,
i.e. vortices\cite{Hove2000,Smiseth2004,Smiseth2005}, since vortices are dual
objects to the currents $\bv b$.  In $3D$, it is known that the dual of a
superfluid is a superconductor,
and vice versa\cite{Kleinert1989,Hove2000,Smiseth2004,Smiseth2005}. Therefore,
the above result for the current-correlator of a superconductor features the same
behavior as the vortex-vortex correlator at a \textit{neutral} fixed point,
since a neutral fixed point in the original theory is a charged fixed point
in the dual theory. Here, $C_2$ is the helicity modulus of the system, $C_3$ is
a critical amplitude, and $C_4$ is essentially the inverse mass of the gauge-field.
The physical interpretation of
$\lim_{\bv q\rightarrow 0}\frac{e^2}{2\beta}\langle \bv{B}_{\bv q}\cdot\bv{B}_{-\bv q}\rangle $
is that when this quantity is zero, there are no long-range correlations of
current-loops in the system, i.e. there are no supercurrents threading the entire
system which is therefore normal metallic. Conversely, when $T < T_c$ this correlator
is non-zero. There are supercurrents threading the entire system, which is therefore
superconducting. When $T>T_C$, the gauge mass will be zero in the long wavelength limit.
When $T<T_C$, however, the factors of $q^2$ will cancel, and the gauge correlator obtains
a finite expectation value, and hence a mass.  The Higgs-mechanism (Meissner effect) in
an $N$-component superconductor is therefore a result of a blowout of closed loops of
charged currents as the temperature is lowered through the phase transition. Conversely,
the transition to the normal state is driven by a collapse of closed current loops, which
is dual to a blowout of closed vortex loops. In either way of looking at the problem, the
Higgs-mechanism is fluctuation driven.

Note that the above result is valid for any number of components $N \geq 1$, and any
value of the  Josephson coupling $\lambda \geq 0$.

The preemptive scenario described in the previous section impacts the temperature-dependence
of the Higgs-mass at the transition from the superconducting to the normal metallic state.
The mass vanishes continuously in an inverted $3DXY$ phase transition if the value of the
gauge-charge is large enough for the preemptive scenario to be ruled out for any $\lambda$,
including $\lambda=0$. For small enough gauge-charge, such that fluctuations in the neutral
sector strongly affect fluctuations in the charged sector, and vice versa, the preemptive
effect comes into play. In that case, the Higgs-mass vanishes discontinuously at the
phase transition.

\section{Conclusion}
\label{sec:conc}

We have formulated an $N$-component London superconductor with intercomponent Josephson
couplings as a model of $N$ integer-current fields $\bv b_\alpha$ and $N(N-1)/2$ monopole fields,
$m_{\alpha,\beta}$. These monopoles allow supercurrents of a particular condensate component
to be converted to a supercurrent of a different component, i.e. currents may change "color"
at any site. For zero Josephson
coupling, $\lambda$, only configurations where all the monopole fields are zero contribute, and
the model reverts to an $N$-component gauge-coupled 3dXY-model. This model is known to have either
i) $N-1$ transitions in the $XY$-universality class and a single phase transition in the inverted
XY-universality class, or ii) a single preemptive first-order phase transition for intermediate
values of the charge. For any $\lambda > 0$, the $N-1$ phase transitions in the neutral sector
are converted to crossovers. In the limit $\lambda\rightarrow\infty$, all orders of monopole
excitations will contribute. This effectively removes the constraints
$\bv\Delta\cdot\bv b_\alpha=0$ on each individual component. There
is only one particular composite mode, $\sum_\alpha\bv b_\alpha$ which is still
divergenceless, and which thus features a phase transition. This transition is
known to be in the inverted 3dXY-universality class for $\lambda=0$. For
small, but finite $\lambda$, fluctuations in the neutral sector are still
substantial although the phase transitions are all converted to  crossovers.
These charge-neutral non-critical fluctuations
nonetheless substantially influence the putative critical fluctuations in the charged
sector, particularly at temperatures close to the $\lambda=0$ $3DXY$ critical
point. This converts the inverted $3DXY$ critical point into a first-order
phase-transition via a preemptive effect. The degree to which the charge-neutral
fluctuations influence the fluctuations in the charged sector for small $\lambda$,
increases with the number of composite charge-neutral fluctuating modes. In the
parameter regime $(e,\lambda)$ where one may have a preemptive effect, the
first-order character of the  superconductor-normal metal phase transition will
therefore be more pronounced with increasing $N$.

As a byproduct of our analysis, we have recast the onset of the photon Higgs-mass
in the superconductor (Meissner effect) in terms of a blowout of current loops
associated with the onset of superconductivity. This analysis goes beyond mean-field
theory and takes all critical fluctuations of the theory into account. The
description giving the onset of the Higgs-mass of the photon in terms of a current-loop
blowout going into the superconducting state as temperature is lowered, is dual to the
description of the vanishing of the Higgs-mass of the photon in terms of vortex-loop
blowout going into the normal metallic state as the temperature is increased.

\begin{acknowledgments}
P.~N.~G. was supported by NTNU and the Research Council of Norway. A.~S. was supported by the Research Council of
Norway, through Grants 205591/V20 and 216700/F20, as well as European Science Foundation COST
Action MPI1201. We tank E. Babaev and J. Garaud for useful discussions.
\end{acknowledgments}

\appendix

\section{The character expansion}
\label{app:charexp}

We apply the expansion
\begin{equation}
  e^{\beta\cos\gamma} = \sum_{b=-\infty}^\infty I_b(\beta)e^{ib\gamma},
\end{equation}
to the cosine terms of \cref{eq:actionlattice}, with $\lambda=0$.
This gives the action
\begin{align}
  \mathcal{Z} ={}&\int\mathcal{D}\bv A\left(\prod_\alpha\int\mathcal{D}\theta_\alpha\right)\nonumber\\
  \times&\prod_{\bv r, \mu, \alpha}\sum_{b_{\bv r, \mu, \alpha}=-\infty}^\infty I_{b_{\bv r, \mu, \alpha}}(\beta)
  e^{ib_{\bv r, \mu, \alpha}(\Delta_\mu\theta_{\bv r, \alpha}-eA_{\bv r,\mu})}\nonumber\\
  \times&\prod_{\bv r}e^{-\frac{\beta}{2}{(\Delta\times\bv A_{\bv r})}^2}
\end{align}
By performing a partial integration of each phase component, $\theta_{\bv r, \alpha}$,  we move the
lattice derivative from the phase to the integer field $\bv b$ in the first term. Then we factorize
the terms dependent on the phases on each lattice site, which may then be integrated separately.
\begin{equation}
  \mathcal{Z}_\theta = \prod_{\bv r, \alpha}\int_0^{2\pi}\text{d}\theta_{\bv r, \alpha} e^{-i\theta_{\bv r, \alpha}(\sum_\mu\Delta_\mu
b_{\bv r, \mu, \alpha})}.
\end{equation}
This constrains the $\bv b$-fields to have zero divergence,
\begin{equation}
  \bv\Delta\cdot\bv b_{\bv r, \alpha} = 0\;\forall\;\bv r, \alpha.
\end{equation}
The partition function then reads
\begin{align}
  \mathcal{Z} = \int\mathcal{D}(\bv A)\sum_{\{\bv b\}}\prod_{\bv r,
  \alpha}\delta_{\bv\Delta\cdot\bv b_{\bv r, \alpha}, 0}\prod_{\bv r, \mu, \alpha}I_{b_{\bv r,
  \mu, \alpha}}(\beta)\nonumber\\
  \prod_{\bv r}e^{-\left[ie\sum_\alpha\bv b_{\bv r, \alpha}\cdot\bv A_{\bv r}+\frac{\beta}{2}{(\bv\Delta\times\bv A_{\bv
 r})}^2\right]}
\end{align}
This represents $N$ integer-current fields which must form closed loops individually, coupled by a single gauge field, $\bv A$.

The next step is to integrate out the gauge degrees of freedom. To this end we Fourier transform the
action
\begin{equation}
  S = \sum_{\bv r}\left[ie\sum_\alpha\bv b_{\bv r, \alpha}\cdot\bv A_{\bv r}+\frac{\beta}{2}{(\bv\Delta\times\bv A_{\bv
r})}^2\right]
 \end{equation}
 into
 \begin{align}
   S = \sum_{\bv q}\Bigg[&\frac{ie}{2}\sum_\alpha\bv b_{\bv q, \alpha}\cdot\bv A_{\bv
 -q}+\frac{ie}{2}\sum_\alpha\bv b_{\bv -q, \alpha}\cdot\bv A_{\bv q}\nonumber\\
 &+\frac{\beta}{2}(\bv Q_{\bv q}\times\bv A_{\bv q})(\bv Q_{-\bv q}\times\bv A_{-\bv q})\Bigg].
 \label{eq:Sq}
\end{align}
Here, we have symmetrized the $\bv b\cdot\bv A$-term, and $\bv Q_{\bv q}$ is the Fourier
representation of the lattice differential operator, $\bv\Delta$. We can further simplify the
expression by choosing the gauge $\bv\Delta\cdot\bv A_{\bv r} = 0$, which translates to $\bv Q_{\bv
q}\cdot\bv A_{\bv q} = 0$ in Fourier space. This reduces the last term to $\beta\left|\bv Q_{\bv
q}\right|^2\bv A_{\bv q}\cdot\bv A_{-\bv q}/2$, where $\left|\bv Q_{\bv q}\right|^2 =
\sum_\mu{(2\sin q_\mu/2)}^2$.
Now we complete the squares in $\bv A_{\bv q}$, to facilitate the Gaussian integration
\begin{align}
  S = \sum_{\bv q}\Bigg[&\Big(\bv A_{\bv q} + \frac{ie}{2}\sum_\alpha\bv b_{\bv q, \alpha}D_{\bv
      q}^{-1}\Big)D_{\bv q}\nonumber\\
      &\times\Big(\bv A_{-\bv q} + \frac{ie}{2}\sum_\alpha\bv b_{-\bv q, \alpha}D_{\bv
      q}^{-1}\Big)\nonumber\\
      &+ \frac{e^2}{4}\Big(\sum_\alpha\bv b_{\bv q, \alpha}\Big)D_{\bv q}^{-1}\Big(\sum_\beta\bv
    b_{-\bv q, \beta}\Big)\Bigg],
\end{align}
where $D_{\bv q} = \beta\left|\bv Q_{\bv q}\right|^2/2$.
Now we can shift and integrate out the gauge field, $A_{\bv q}$, which leaves us with
\begin{equation}
  S = \sum_{\bv q}\frac{e^2}{2\beta\left|\bv Q_{\bv q}\right|^2}\Big(\sum_\alpha\bv b_{\bv q,
  \alpha}\Big)\cdot\Big(\sum_\beta\bv b_{-\bv q, \beta}\Big),
  \label{eq:Cintq}
\end{equation}
or in real space
\begin{equation}
  S = \sum_{\bv r, \bv r^\prime}\frac{e^2}{2\beta}\Big(\sum_\alpha\bv b_{\bv r,
  \alpha}\Big)\cdot\Big(\sum_\beta\bv b_{\bv r^\prime, \beta}\Big)D(\bv r - \bv r^\prime).
  \label{eq:Cintr}
\end{equation}
Here, $D(\bv r-\bv r^\prime)$ is the Fourier transform of $1/\left|\bv Q_{\bv q}\right|^2$.
Inserting this into the action, we arrive at
\begin{align}
  \mathcal{Z} = \sum_{\{\bv b, m\}}&\prod_{\bv r, \alpha}\delta_{\bv\Delta\cdot\bv b_{\bv r, \alpha}, 0}
\prod_{\bv r, \mu, \alpha}I_{b_{\bv r, \alpha, \mu}}(\beta)\nonumber\\
&\prod_{\bv r, \bv r^\prime}e^{-\frac{e^2}{2\beta}\sum_{\alpha, \beta}\bv b_{\bv r, \alpha}\cdot\bv
b_{\bv r^\prime,\beta}D(\bv r - \bv r^\prime)},
\end{align}
which is \cref{eq:besselJ0}

\section{Two-dimensional multiband superconductors}
\label{app:2D}

In a thin-film superconductor, the effective magnetic penetration depth is inversely proportional to
the film thickness. Hence, in a two-dimensional system, the magnetic penetration depth becomes
infinite, and the effective charge of the charge carriers become zero. This effectively freezes out
the gauge-field fluctuations of the interior of the film, in turn eliminating the long-range
gauge-field mediated vortex-vortex interactions. In this case the relevant lattice action will be

\begin{align}
  S = -&\beta\sum_{\bv r}\sum_{\mu, \alpha}\cos\left(\Delta_\mu\theta_{\bv r,
  \alpha}\right)\nonumber\\
  -&\beta\lambda\sum_{\bv r}\sum_{\alpha<\beta}\cos\left(\theta_{\bv r, \alpha}-\theta_{\bv r, \beta}\right).
  \label{eq:2dactionlattice}
\end{align}
That is, it is effectively a neutral condensate.

We may apply the character expansion of \cref{eq:besselid} to \cref{eq:2dactionlattice}, which
results in the partition function
\begin{align}
  \mathcal{Z} = \sum_{\{\bv b, m\}}&\prod_{\bv r, \alpha}\delta_{\bv\Delta\cdot\bv b_{\bv r, \alpha},
\sum_{\beta\neq\alpha}m_{\bv r, \alpha, \beta}}\nonumber\\
&\prod_{\bv r, \mu, \alpha}I_{b_{\bv r, \alpha, \mu}}(\beta)  \prod_{\bv r, \alpha<\beta}I_{m_{\bv
r, \alpha, \beta}}(\beta\lambda).
\label{eq:partbJ2d}
\end{align}
This is of course very similar to \cref{eq:partbJ}, with the differences being as follows. The
integer-current field, $\bv b_{\bv r}$ is now a two-component vector, as is naturally the position
vector, $\bv r$, and the gauge-field mediated interaction has disappeared.

We may apply the same reasoning to \cref{eq:partbJ2d} as we did in the main text. There will be a
single mode, $\sum_\alpha\bv b_{\bv r}$ which is divergenceless, and $N(N-1)/2$ modes with finite
divergence. The only difference now in the two-dimensional case is the lack of gauge-field mediated
interactions in the divergenceless mode. Hence, the single remaining phase transition is expected to
be a Kosterlitz-Thouless transition from a two-dimensional superfluid to a normal fluid. This
prediction could be verified in Monte-Carlo simulations, as the partition function of
\cref{eq:partbJ2d} is particularly well suited for worm-type algorithms.

\bibliography{ref}
\end{document}